\documentclass[aps,showpacs,onecolumn,floats,prd,superscriptaddress,nofootinbib]{revtex4-1} 
\usepackage{graphicx,amsmath,amssymb,amstext}
\usepackage{amssymb,amsbsy,amsfonts,amsthm,color}
\usepackage{epsfig}
\usepackage{graphicx}
\usepackage{subfigure}
\graphicspath{{Figures/}}

\begin{document}

\title{Memory effect from supernova neutrino shells}

\author{Darsh Kodwani}
\email{dkodwani@physics.utoronto.ca}
\affiliation{Canadian Institute of Theoretical Astrophysics, 60 St George St, Toronto, ON M5S 3H8, Canada.}
\affiliation{University of Toronto, Department of Physics, 60 St George St, Toronto, ON M5S 3H8, Canada.}

\author{Ue-Li Pen}
\email{pen@cita.utoronto.ca}
\affiliation{Canadian Institute of Theoretical Astrophysics, 60 St George St, Toronto, ON M5S 3H8, Canada.}
\affiliation{Canadian Institute for Advanced Research, CIFAR program in Gravitation and Cosmology.}
\affiliation{Dunlap Institute for Astronomy \& Astrophysics, University of Toronto, AB 120-50 St. George Street, Toronto, ON M5S 3H4, Canada.}
\affiliation{Perimeter Institute of Theoretical Physics, 31 Caroline Street North, Waterloo, ON N2L 2Y5, Canada.}

\author{I-Sheng Yang}
\email{isheng.yang@gmail.com}
\affiliation{Canadian Institute of Theoretical Astrophysics, 60 St George St, Toronto, ON M5S 3H8, Canada.}
\affiliation{Perimeter Institute of Theoretical Physics, 31 Caroline Street North, Waterloo, ON N2L 2Y5, Canada.}

\begin{abstract}
When a supernova explodes, most of its energy is released in a shell of relativistic neutrinos which changes the surrounding geometry.
We calculate the potentially observable responses to such a change in both pulsar scintillation and conventional interferometers.
In both cases, the responses are permanent changes due to such a transient event.
This is by-definition a memory effect.
In addition to the transverse component in the usual gravitational memory (Christodolou effect) effect, it also has a longitudinal component. 
Furthermore it is different from the Christodolou effect as the transverse component of this memory effect also has a term that grows with time. 
\end{abstract}

\maketitle

\section{Introduction and Summary}

In recent papers \cite{Olum:2013gza, PhysRevD.93.103006}, we and Olum et al have calculated the effect a shell of matter can have on photon geodesics when it crosses them. 
In particular the effect is calculated for photons coming from a pulsar, and the shell of matter is made of relativistic neutrinos coming from a nearby supernova.
The change in the local geometry alters the path of photons.
Observationally, this manifests itself as a permanent shift in pulsar signal arrival times. 

In this paper, we extend the above scenario to multiple photon trajectories.
This situation arises naturally in pulsar scintillation, which is caused by the photons propagating to us along multiple paths. 
Each path will be affected by the same effect that shifts their individual arrival times with slightly different magnitudes.
This ``relative'' time shift between two paths is much less than the absolute time shift calculated in \cite{Olum:2013gza, PhysRevD.93.103006}. 
Nevertheless, one can measure it as an interference effect.
The accuracy is then determined by the signal frequency, which can be much better than pulse-profile timing \cite{PenYan14}.
We provide a simple estimation to show that among the thousands of pulsars that SKA is supposed to discover \cite{MSPpopulation}, some of their lines-of-sight can be sufficiently close ($\sim 10$ light-years) to the next supernova.
If these pulsars do scintillate, the supernova causes an order one change in their scintillation patterns over the course of $\sim$ 10 years. 

Since the passage of a massive shell causes a permanent time shift between two nearby null trajectories, it is by-definition a memory effect.
We proceed to clarify its relation and difference to the gravitational memory effect \cite{Christodoulou_effect,GW_memory} by computing the direct response from an interferometer.

\begin{itemize}
\item It has a longitudinal component. The transverse-traceless limitation only applies to freely-propagating changes of the metric (i.e gravitational waves). 
While coupled to matter, which often have longitudinal (density) waves, it is natural to have an accompanying longitudinal change in the metric. 
The longitudinal component will be a permanent shift in the distance between two geodesics, therefore a permanent change in the arm-length of an interferometer.
\item In addition to displacements of geodesics, the transverse component will also have a term that grows with time. 
In other words, two geodesics which were at rest suddenly pick up a relative velocity toward each other.
\end{itemize}
In terms of the dynamics, a change in velocity is higher order than a change in distance.
This however does not mean that our effect is harder to measure. 
The conventional gravitational memory effect needs a physical event that significantly breaks spherical symmetry. 
In contrast, our effect does not, so it can occur more generally.\footnote{Note that this is different to the effect calculated by Epstein \cite{Epstein:1978dv} as that also required a breaking of spherical symmetry.}
In addition, a change in velocity implies a distance change that grows in time even after the initial effect. 
That is an advantage for some detection methods.  

Unfortunately, as long as the supernova is reasonably far away, this effect is too weak for existing or future interferometers. 
Pulsar scintillometry seems to be the most likely method to make the first detection. 
Such a detection will allow us to derive the energy emitted by a supernova in neutrinos. 
Currently there is no other way to obtain such information. 
Thus in addition to direct neutrino detection experiments such as in Super-Kamiokande \cite{SuperKSN}, this memory effect can provide an independent constraint on the explosion mechanism.

The rest of the paper is organized as follows. 
In section \ref{sec-scint} we calculate how a pulsar scintillation pattern can be sensitive to the passage of a neutrino shell.
We put in reasonable numbers to show that an observation is likely to happen if SKA discovers thousands of pulsars as expected.
In section \ref{RelV} we derive the interferometer response in both transverse and longitudinal directions.  
We put in reasonable values to estimate the size of the interferometer response.
Unfortunately the amplitude of the effect is well below the sensitivity of the experiments that are currently being planned such as LISA and BBO.
Thus pulsar scintillometry remains to be the most likely method of detection in the near future.

\section{Memory effect in Pulsar Scintillation}
\label{sec-scint}

\begin{figure}[t!]
\begin{center}
\includegraphics[width=\textwidth,height=10cm]{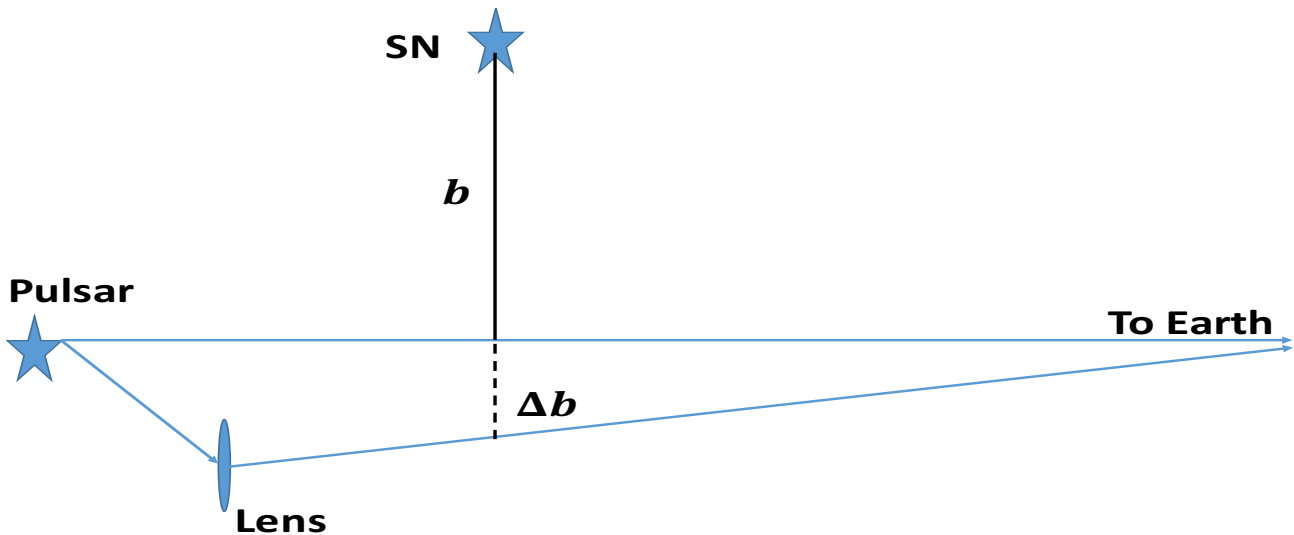}
\caption{Geometry of the astrophysical interferometer formed by pulsar scintillometry. Due to scattering or lensing, the image we see is an interference pattern of two light rays represented by the blue lines. If the separation of the two light rays has a component along the longitudinal (radial) direction from the SN, the spacetime distortion of the neutrino shell will change the interference pattern we see. We draw the lens to be behind the SN, but it could have been in front of it and the effect is the same.}
\label{fig:4}
\end{center}
\end{figure}

It is known that the images of many astronomical bodies scintillate \cite{PulsarScint}. A general reason for scintillation is that due to scattering or lensing, we receive multiple light rays from the same objects. These light rays are very close to each other, so they cannot be individually resolved and have to interfere. The scintillation pattern we see is the time dependence of their interference. If we consider two light rays from a faraway pulsar which happen to pass by a SN progenitor, as illustrated in Fig.\ref{fig:4}, they can probe the spacetime distortion when it explodes. 

The scintillation/interference pattern is directly related to the path lengths of these light rays. The change in such path length during a SN explosion has been worked out in \cite{Olum:2013gza}
\begin{equation}
	\Delta t = 2\delta M \left[ \ln \left(1 + \frac{t^2}{b^2} \right) - \frac{t^2}{b^2 + t^2} \right].
\end{equation}
Here $b$ is the impact parameter as shown in Fig.\ref{fig:4}, the shortest distance between the light ray and the SN. $t$ is the proper time on earth, with $t=0$ the time we directly observe the SN explosion. $\delta M$ is the total energy of the neutrino shell, and $\Delta t$ is the resulting time shift. A photon which should have reached the earth at time $t$, will arrive earlier at $(t-\Delta t)$ instead.

When the separation between two light rays has a component in the radial direction from the SN, $\Delta b$, there will be a nonzero relative change between their path lengths.
\begin{equation}
	(\Delta t|_b - \Delta t|_{b+\Delta b}) \approx 
	\frac{\partial \Delta t}{\partial b} \Delta b 
	= - \frac{4\delta Mt^4}{b(b^2 + t^2)^2} \Delta b~.
	\label{eq-change}
\end{equation}
We can see that this effect grows from zero and approaches an asymptotic value,
\begin{equation}
	(\Delta t|_b - \Delta t|_{b+\Delta b}) 
	\longrightarrow \frac{4\delta M \Delta b}{b}~,	\label{pulsescint}
\end{equation}
at a characteristic time scale given by $b$.

We estimate $b$ by assuming that the next SN is somewhere near the galactic centre. 
A sample of $\sim 9000$ pulsars from the SKA catalog in \cite{MSPpopulation} shows that among those pulsars, the shortest $b$ is about $10\ ly \sim 10^{14} \ km$. 
$\Delta b$ is related to the scattering-broadening of images. 
We use the data from \cite{BowBel13} that was observed on a scattering screen near the galactic centre. 
Scaling the frequency to $1\ GHz$ which is usually a good window to observe pulsar signals. 
We found that such a scattering screen can produce images separated by $\Delta b\sim 1000A.U. \sim 10^{10} \ km$. 
We again use $\delta M \sim 1 \ km$, and combining all these numbers we get $(\delta M \Delta b / b) \sim 1 \ m$. 
This is comparable to the wavelength at $1 \ GHz$, which means that an interference bright spot becomes a dark spot within the time scale set by $b\sim$ 10 years. 
As long as the original scintillation pattern is observable, such an order-one change is guaranteed to be observable without a detailed analysis of $S/N$.

The actual observational concern is that the scintillation pattern can change during those 10 years without the influence from a near-by supernova.
Such change can be caused by both the relative motion between the pulsar and the screen, and the internal dynamics of the screen.
For 10 years or even a longer time scale, the relative motion is usually well-approximated as a constant velocity.
Thus, if we monitor the scintillation pattern for 10 more years both before and after the supernova event, we will be able to fit this background change and remove it to isolate the supernova contribution.
As for the internal dynamics of the screen, if it also changes at a natural time scale $\gtrsim$ 10 years, then it will be removed together by the above procedure.
If the dynamical time scale is comparable or shorter than 10 years, then we need to understand it better to determine how much of the observable change can be attributed to the supernova.
The intrinsic dynamics of the scattering screen is an active research topic \cite{Liu:2015, Pen:2013} (and references there in), so we do not know which one is the case yet.

\section{Memory effect in interferometers}
\label{RelV}

\begin{figure}[h!]
\begin{center}
\includegraphics[scale = 0.27]{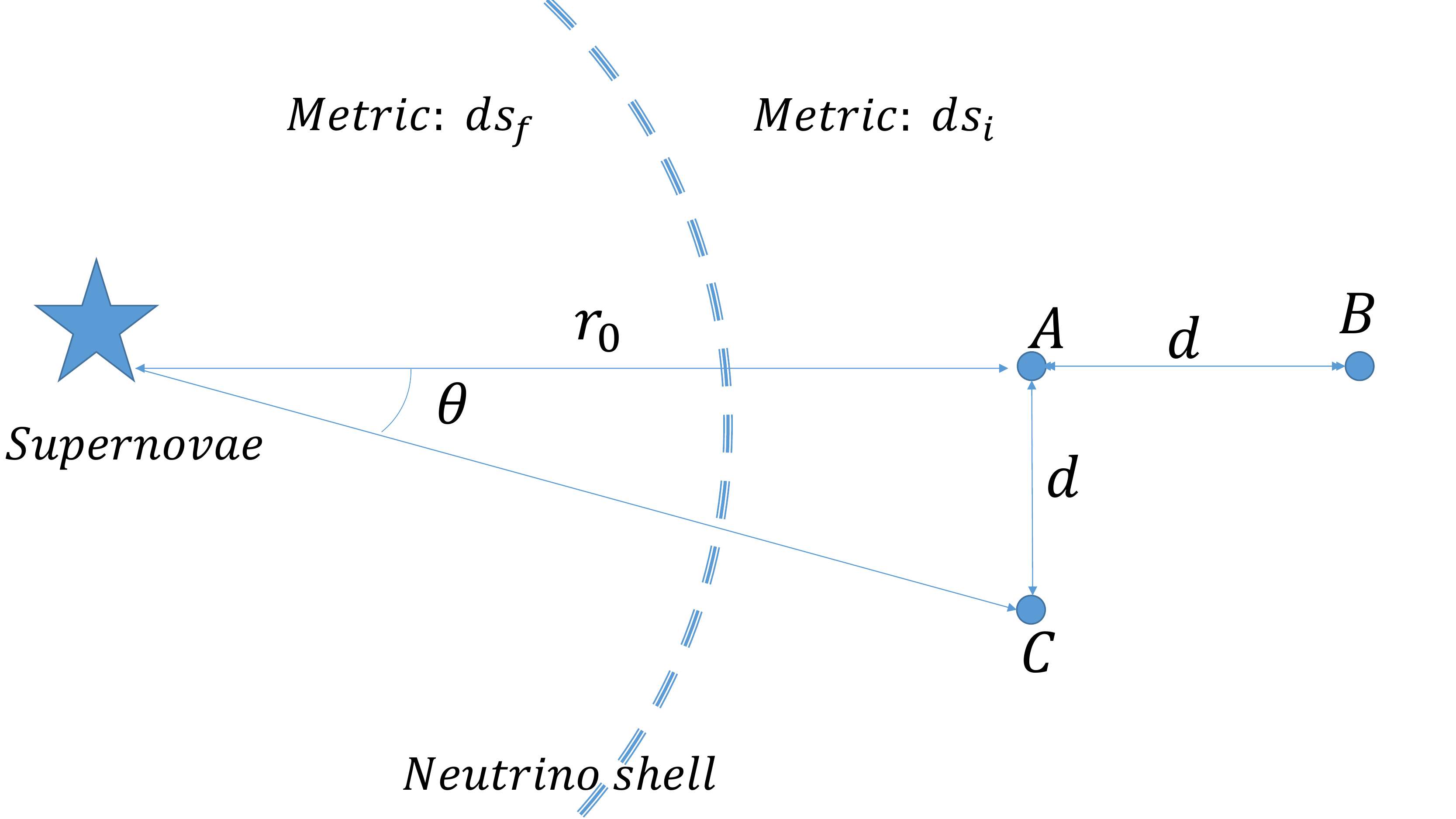}
\caption{Schematic of the effect being considered by a neutrino shell passing through the interferometer. The points $A,B,C$ represent ends of the interferometer of arm length $d$. The three points $A,B$ and $C$ will pick up velocities $v_A, v_B$ and $v_C$ respectively after the shell crosses them. They are all different since they cross the shell at different locations.}
\label{fig:1}
\end{center}
\end{figure}

The change in the scintillation pattern is an integrated effect along the line-of-sight. 
Obviously, most of this effect comes from the region closest to the supernova.
Here we wish to understand the details of the geometric change in that region.
We will imagine that there is actually an interferometer somewhere near the supernova, as shown in Fig.\ref{fig:1}.
We assume the 3 end-points of the 2 arms of this interferometer follow geodesics.
From the response of this gedankin interferometer, we can better understand how the local geometry is affected by the passage of the shell.

The immediate effect of a geodesic crossing the shell is an apparent velocity change.
For a geodesic at rest with respect to the supernova immediately before shell-crossing, it picks up a radial velocity with respect to the remnant immediately after shell crossing.
We calculate this velocity using the Israel junction conditions in Appendix \ref{sec-junc}.
Using the asymptotic proper time $t$ at rest with the remnant, the velocity is given by.
\begin{eqnarray}
	\frac{dr}{dt} = -\frac{\delta M}{r_{crossing}} \left( 1 + \frac{\delta M}{r_{crossing}} \right),
	\label{eq-dv}
\end{eqnarray}
where $r_{crossing}$ is the radial coordinate at which the shell crosses the geodesic.

We have kept 1 subleading order in this result.
Even the leading order effect is beyond Newtonian gravity, since there is actually no momentum transfer between he shell and the geodesic.
In fact, the geodesic picks up an infalling velocity, which is opposite to any intuition about momentum transfer.
This is purely a relativistic effect that the rest-frame of the supernova progenitor and the rest-frame of the remnant are different.
The explicit relativistic calculation will be given in Appendix \ref{sec-junc}.

If we take the initial velocity as an input, its effect on the gedankin interferometer actually follows simple Newtonian intuitions.
Let us first consider the interferometer arm between $A$ and $B$, whose length we denote as $l_{AB}$.
From Eq.~(\ref{eq-dv}), one might expect an asymptotic velocity difference between $A$ and $B$, which allows $l_{AB}$ to change with time even after the shell crosses.
However, if we recall that the remnant provides a weaker acceleration, and point $A$ starts to suffer from that before point $B$, such difference exactly cancels the velocity difference from the shell.
Up to leading order, there is only a constant change to $l_{AB}$, coming from the fact that an instant velocity change and a gradual change due to acceleration cannot cover the same distance traveled.
\begin{equation}
	\Delta l_{AB} \equiv l_{AB} - \bar{l}_{AB} = - \frac{\delta M^2 d}{r_0^2}.
	\label{eq-LAB}
\end{equation}
An explicit relativistic calculation is given in Appendix \ref{sec-LAB}.
Strictly speaking, an interferometer measures the proper time for a photon to propagate back and forth between $A$ and $B$.
It turns out that at the leading order, $l_{AB}$ is half the proper time in the rest frame of the SN progenitor (or remnant). This is confirmed in Appendix \ref{sec-ptpl}.

On the other hand, points $A$ and $C$ get the same speed, but their velocities are at a slightly different angle.
It is a simple trigonometry calculation (given in Appendix \ref{sec-lAC}) to show that they acquire a velocity towards each other right after shell crossing.
\begin{equation}
	\Delta l_{AC} \equiv l_{AC} - \bar{l}_{AC} = - \frac{\delta M d}{r_0^2} t.
	\label{eq-LAC}
\end{equation}
At the time-scale longer than $10^{-5}$ seconds (the Schwarschild radius for $\delta M\sim$ solar-mass), the change in $l_{AC}$ is the dominant response from this gedanken interferometer.

From Eq.~(\ref{eq-LAB}) and (\ref{eq-LAC}), we can see that the gedankin interferometer gets an anisotropic stretching, which is similar to a time-dependent strain.
\begin{equation}
h \sim \frac{\Delta L}{d} \sim \frac{\delta M t}{r_0^2}~.
\label{eq-strain}
\end{equation} 
We can use this to estimate whether an interferometer may detect such a signal.

Since our strain grows linearly with time, we do not expect detections from ground based experiments as for those setups the three points $A,B,C$ cannot remain in free fall for a long enough amount of time such that the signal builds up to an observable value. 
If we plot our effect on the strain-frequency diagram \cite{GWcurves} that is usually used to compare different interferometers, it will be a 45-degree line. Thus the first point at which the sensitivity curve of a device crosses with a 45-degree line will give the best chance for our effect being detected. In all these estimations, we take $\delta M$ to be a faction of a solar mass, and take the corresponding Schwarzschild radius to be $1~km$ for simplicity.

For LISA, the best observing frequency is $\sim 0.5 \times10^{-2} Hz$ with a sensitivity in strain $\sim 10^{-21}$. Using Eq.~(\ref{eq-strain}), we can solve for the distance to the SN, $r_0$, for our effect to be detectable.
\begin{eqnarray}
	r_0 & = &  \left(  \frac{\delta M}{h} t \right)^\frac{1}{2} \label{Meas}	\\
	& = &  \left( \frac{1 \ km}{10^{-21}} \times 10^7 \ km \right)^\frac{1}{2} \approx 10^{14} \ km = 10 \ ly.
\end{eqnarray}
This is clearly too close. It has been estimated that only once in $10^8$ years will a SN go off within a distance of $30 \ ly$ \cite{EllSch93}.\footnote{And if that happens, it might kill us.} By a naive volume scaling, an explosion within 10 $ly$ only occurs once every billion years.

If we look at the Big Bang Observer (BBO) instead, the best observing frequency is $\sim 0.5 Hz$ with a sensitivity in strain $\sim 10^{-24}$. First of all, this frequency range does not have as many background signals from compact binaries, making it a much better device to measure our effect. The improved sensitivity gives a value for $r_0$ of $\sim 100 \ ly$. This is a factor of $10^3$ increase in the volume for detectable events, thus improves the expectation of one SN that is within $10 \ ly$ to occur in less than a million years . That is unfortunately still a long shot.

In this type of simple estimation, we cannot go lower in the frequency. The exact duration of the neutrino-shell passage is not known, but we do no expect it to be much less than a second. Thus for higher frequencies, the co-dimension-one delta function approximation breaks down, and the effect will be weaker than Eq.~(\ref{eq-strain}).

Finally, we expect 2 to 3 SN per century in our galaxy and we can assume that the next SN would be at a distance comparable to the galactic diameter of $\sim 10^5 \ ly$. If we are going to measure such an effect at $1 \ Hz$, again using Eq.~(\ref{eq-strain}), we find
\begin{equation}
	h = \frac{ 1~km \times ~ (3\times10^8~m)}{(10^5~ly)^2}\sim 10^{-30}~.
\end{equation}
This requires a measurement of the strain that is six orders of magnitude better than BBO and is not yet achievable by interferometers that are currently being planned.

\acknowledgments

This work is supported by the Canadian Government through the Canadian Institute for Advance Research and Industry Canada, and by Province of Ontario through the Ministry of Research and Innovation.

\appendix
%%%%%%%%%%%%%%%%

\section{Apparent velocity change from junction conditions}
\label{sec-junc}

We will use Schwarzschild metric to describe the geometry around the SN progenitor of mass $M$.
When it releases a spherically symmetric shell of neutrinos, the metric outside remains the same.
The metric inside becomes a different Schwarzschild metric with mass $(M-\delta M)$, where $\delta M$ will be the energy carried by the neutrinos after the SN explosion. 
The physical effect we will calculate is due to crossing this shell.
Contribution to the metric from other sources can be linearly superimposed and will not change our final result, so we can ignore them.
The outside metric will have a bar over its coordinates and inside one will not. 
 
\begin{equation}
	d\bar{s}^2 = \bar{g}_{\mu \nu} d\bar{x}^\mu d\bar{x}^\nu = - \left( 1 - \frac{2M}{r} \right) d\bar{t}^2 + \left( 1 - \frac{2M}{r} \right)^{-1} d {r}^2 + r^2 d {\Omega}_2^2. \label{SCH}
\end{equation}

\begin{equation}
	ds^2 = g_{\mu \nu} dx^\mu dx^\nu = - \left( 1 - \frac{2(M - \delta M)}{r} \right) dt^2 + \left( 1 - \frac{2(M - \delta M)}{r} \right)^{-1} dr^2 + r^2 d \Omega_2^2. \label{SCHa}
\end{equation}
We are working in units with $G = c =1$.  Notice that the time component of the metric is different in both geometries whereas the radial component is the same as it corresponds to the radius of a two sphere separating the two geometries.  We describe the shell as a delta function travelling at roughly the speed of light and thus following a null geodesic. The null vector of the shell can be written in both metrics as follows\footnote{One could write this vector in different ways and still have it satisfy the null normalization condition. However the vectors should have the same radial component, since the $r$ coordinate in both metrics are identified by the physical area.}
\begin{equation}
	k^\mu = \left( -g_{tt}^{-1}, 1, 0, 0 \right), \hspace{5mm} \bar{k}^\mu = \left( -\bar{g}_{tt}^{-1}, 1, 0, 0 \right)	\label{sigma}
\end{equation}
Since we expect the interferometer points (denoted by the points $A,B,C$ in figure \ref{fig:1}) to be very far from the massive object we assume they have a negligible initial velocity.
\begin{equation}
	\bar{u}^\mu = \left( (-1/\bar{g}_{tt})^\frac{1}{2}, 0, 0, 0 \right).	\label{zeta}
\end{equation}
Assuming that the material of the interferometer does not directly interact with neutrinos, its motion does not change while crossing the junction. 
That means the inner product between the geodesic and the shell remains the same.\begin{equation}
	 g_{\mu \nu} u^\mu k^\nu = \bar{g}_{\mu \nu} \bar{u}^\mu \bar{k}^\nu. 	\label{0IJC}
\end{equation}
In order to satisfy this relation, this geodesic needs to pick up a velocity $v$ when the shell crosses it.
It is not because its velocity has changed due to any momentum transfer.
It is reflecting the fact that the ``rest frame'' of the two metrics do not agree with each other.
Someone at rest in the first metric has a nonzero velocity in the second metric.

Throughout this calculation we assume that the interferometer is very far from the supernova and the interferometers' arm length is much smaller then the distance to the supernova. Thus we assume there are three small quantities, $\frac{M}{r_0}, \frac{\delta M}{r_0}, \frac{d}{r_0}\ll1$. To simplify notation we will use the symbol $\mathcal{O}(r^{-1}_{0})$ to represent suppression by any one of the three small quantities. 
Substituting Eqs (\ref{sigma}, \ref{zeta}) into Eq (\ref{0IJC}) gives the final vector for the interferometer to order $\mathcal{O}(r_{crossing}^{-2})$,

\begin{equation}
	u^\mu = \left( (-1/g_{tt})^\frac{1}{2} \sqrt{1+v^2g_{rr} }, v , 0, 0 \right).	\label{5}
\end{equation}
The change in velocity due a shell crossing is $v = -\frac{\delta M}{r_{crossing}} \bar{g}_{tt}^{-\frac{1}{2}} $ . $r_{crossing}$ is a fixed distance at which the shell crosses a point. For $A$ it is $r_0$, for $B$ it is $r_0 + d$ and for $C$ it is $r_0 + \mathcal{O}(r_0^{-1})$. Note that this is the proper velocity $\frac{dr}{d\tau}$. We will need the coordinate velocity for our calculations which is given by 

\begin{eqnarray}
	\frac{dr}{dt} &=& \frac{dr}{d \tau} \frac{d \tau}{dt} = v g_{tt}^{\frac{1}{2}} = - \frac{\delta M}{r_{crossing}} \bar{g}_{tt}^{-\frac{1}{2}} g_{tt}^\frac{1}{2} \nonumber \\
	&=& -\frac{\delta M}{r_{crossing}} \left( 1 + \frac{\delta M}{r_{crossing}} \right).
\end{eqnarray}
We will use this coordinate velocity to define the trajectory in the following appendices. 

%%%%%%%%%%%%%%%%%%%%%%%%%

\section{Proper length calculations}
\subsection{Proper length change between $A$ and $B$}
\label{sec-LAB}

\begin{figure}[tb!]
\begin{center}
\includegraphics[width =\textwidth, height = 12cm]{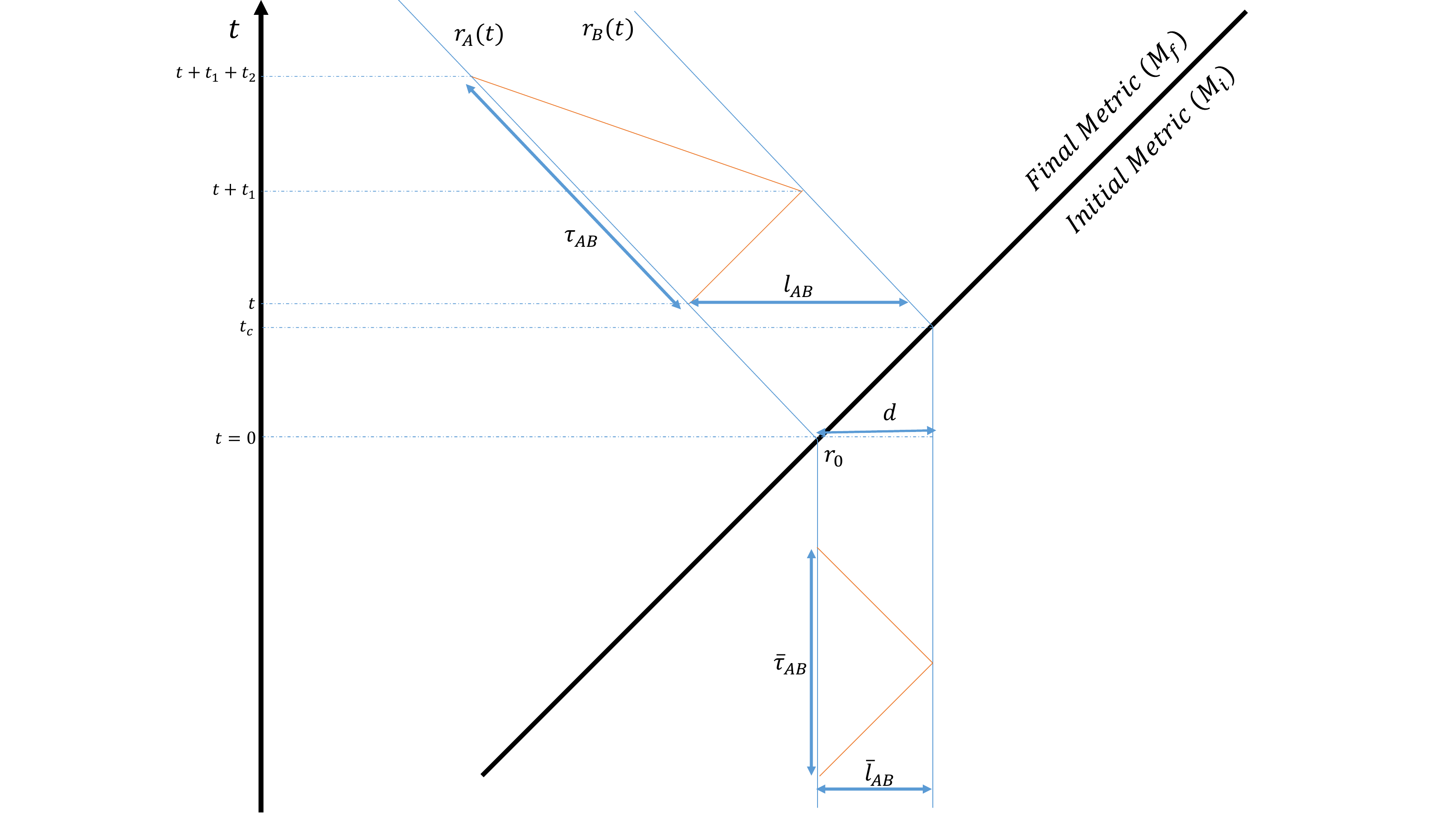}
\caption{Spacetime diagram showing the paths of photons that are used in the interferometer to measure the change in the length of the interferometer arms. The orange lines represent the photon trajectories. The solid blue lines represent the trajectories of the two points $A$ and $B$ in figure \ref{fig:1}. The dark black line is the null trajectory of the neutrino shell. The proper lengths before(after) shell crossing are $\bar{l}_{AB}(l_{AB})$. Proper times before(after) shell crossing are $\bar{\tau}_{AB}(\tau_{AB})$. The coordinate time at which the shell crosses point $B$ is different in both metrics. $t_c$ in metric of mass $M-\delta M$ and $\bar{t}_c$ in the metric of mass $M$. }
\label{calculation}
\end{center}
\end{figure}

Points $A$ and $B$ are the end points of an interferometer arm that is radially oriented towards the SN (as shown in figure \ref{fig:1}). 
Here we calculate the interferometer response in this arm.
As shown in figure \ref{calculation}, such response is actually about the proper time in one end when a signal bounces back.
However, we will show that to the order we care about, such proper time is just the proper length of the arm in the global rest frame.
Therefore we will simply calculate such a proper length.

\subsubsection{Defining trajectories}

We parametrize the trajectories of the points in a piecewise form. 
The trajectory for point $A$ is
\begin{eqnarray}
	r_A(\bar{t}) & = & r_0 - \underbrace{\frac{1}{2} \bar{a}_{(2)}^{(A)} \bar{t}^2}_{T1}, \hspace{5mm} (\bar{t} \leq 0)	\nonumber	\\
	r_A(t) & = & r_0 - \underbrace{\frac{1}{2} a^{(A)}_{(2)} t^2}_{T2} - \underbrace{\left(v^{(A)}_{(1)} + v^{(A)}_{(2)} \right)t}_{T3},  \hspace{5mm} (t>0)  \label{rA}.
\end{eqnarray}

We have introduced new notation to represent the velocity and acceleration of the particles. The subscript represents the number of factors of $r_0$ the term is suppressed by, i.e $a^{(A)}_{(2)}$ is term of $\mathcal{O}(r_0^{-2})$.\footnote{We will only keep terms up to order $\mathcal{O}(r_0^{-2})$.} The superscript is a label for the particle. Here $\bar{a}^{(A)}_{(2)}$ is the acceleration of particle $A$ in the initial metric. $a^{(A)}_{(2)}$ is the acceleration of particle $A$ in the final metric. $v^{(A)}_{(1)}(v^{(A)}_{(2)})$ represent the first(second) order velocities of particle $A$ in the final metric. 
\begin{itemize}
\item $t=0$ is defined to be the time when the shell crosses point $A$ at $r_A(0) = r_0$. For $t<0$ we assume there is no velocity. The only term that contributes to the trajectory is the acceleration of point $A$, $\bar{a}^{(A)}_{(2)}$ towards mass $M$. This is given by $T1$.
\item For $t>0$ there is still the acceleration term, $a^{(A)}_{(2)}$, but the mass is different since the shell has carried $\delta M$ away. This is given by $T2$.
\item $T3$ comes from the velocity that is picked up by shell crossing. 
\end{itemize}

These accelerations and velocities take the following values.
\begin{equation}
	\bar{a}^{(A)}_{(2)} = \frac{M}{r_0^2}, \ \ \ 
	a^{(A)}_{(2)} = \frac{M - \delta M}{r_0^2}, \ \ \ 
	v^{(A)}_{(1)} = \frac{\delta M}{r_0}, \ \ \ 
	v^{(A)}_{(2)} = \frac{\delta M^2}{r_0^2}.
\end{equation} 

The trajectory of point $B$ is given by
\begin{eqnarray}
	r_B(\bar{t}) & = &  r_0 + d  - \underbrace{\frac{1}{2} \bar{a}_{(2)}^{(B)} \bar{t}^2}_{T4},  \hspace{5mm} (\bar{t} \leq \bar{t_c})	\nonumber	\\
	r_B(t) & = & r_0 +d  - \underbrace{\frac{1}{2} a_{(2)}^{(B)}(t-t_c)^2}_{T5} -  \underbrace{\frac{1}{2}  \bar{a}^{(B)}_{(2)} \bar{t}_c^2}_{T6} - \underbrace{\left( \bar{a}_{(2)}^{(B)} \bar{t}_c + v^{(B)}_{(1)} + v^{(B)}_{(2)} \right)(t-t_c)}_{T7} ,  \hspace{5mm}  (t>t_c).	\label{rB}
\end{eqnarray}
The accelerations and velocities are defined analogously to point $A$. 
\begin{itemize}
\item Note that the time coordinates do no agree in the two metrics. 
We have already used the shift symmetry to demand that $t=\bar{t}=0$ is when point $A$ crosses the shell.
The time for point $B$ to cross the shell, $t_c$ and $\bar{t}_c$, must be different.
\footnote{Since the only thing the point does in the initial metric is accelerate, $\bar{t}_c$ will always multiply an acceleration term. 
We will only be interested in terms of $\mathcal{O}(r_0^{-2})$ and the acceleration term will already be $\mathcal{O}(r_0^{-2})$. 
This means throughout our calculation we will never need to distinguish between $\bar{t}_c$ and $t_c$ as they are both $d$ to leading order.}
\item For $\bar{t} <\bar{t}_c$ there is just the acceleration, $\bar{a}^{(B)}_{(2)}$, of the point towards mass $M$. This is given by $T4$.  
\item After $t>t_c$ the shell has crossed. The particle now accelerates in the final metric.
Term $T5$ describes the influence of the new acceleration. 
\item $T6$ corresponds to the distance travelled while accelerating in the initial metric for time $t_c$. 
\item $T7$ is a term that contains the velocity of the particle right after crossing. The $\bar{a}^{(B)}_{(2)} t_c$ term is the velocity gained while accelerating in the metric of mass $M$. $v_{(1)}^{(B)}, v_{(2)}^{(B)}$ are the velocities picked up due to shell crossing.
\end{itemize}
The values of these 1st and 2nd order accelerations and velocities are summarized here.
We repeat the values for point $A$ to show that some of them are actually the same.
\begin{eqnarray}
	\bar{a}^{(A)}_{(2)} & = & \bar{a}^{(B)}_{(2)} =  \frac{M}{r_0^2}	\nonumber	\\
	a^{(A)}_{(2)} & = & a^{(B)}_{(2)}=  \frac{M - \delta M}{r_0^2}	\nonumber	\\
	v^{(A)}_{(1)} & = & v^{(B)}_{(1)} =  \frac{\delta M}{r_0} 	\nonumber	\\
	v^{(A)}_{(2)} & = &  \frac{\delta M^2}{r_0^2}		\nonumber	\\
	v^{(B)}_{(2)} & = & \frac{\delta M^2}{r_0^2} - \frac{\delta M d}{r_0^2}.		
\end{eqnarray}

\subsubsection{Relation between proper length and proper time}
\label{sec-ptpl}
The interferometer infers the distance between two points by measuring the proper time it takes for a photon to go from one end of the arm to the other and come back. As promised we will now relate the proper time seen by one end of the arm to the proper length of that arm (see figure \ref{calculation}). 
In particular we show that \emph{the proper time is twice the proper length at a fixed coordinate time up to $\mathcal{O}(r_0^{-2})$}. This will be very useful as we can compute the proper length (which is comparatively easier) and still infer the effect seen in the interferometer.

 The most general expression for the proper distance is 

\begin{equation}
	l = \int \sqrt{g_{ab} dx^a dx^b} \hspace{5mm} a,b \in \{r,\theta,\phi\}.	\label{proplen}
\end{equation}
Using this we calculate the proper distance before shell crossing $\bar{l}_{AB}$ between the two points $A$ and $B$ at time $\bar{t}$. Since we are only considering radial motion Eq (\ref{proplen}) simplifies to

\begin{equation}
	\bar{l}_{AB} = \int^{r_B(\bar{t})}_{r_A(\bar{t})} dr \ \bar{g}_{rr}^\frac{1}{2}  = \int^{r_B(\bar{t})}_{r_A(\bar{t})} dr \left( 1 + \frac{M}{r} + \frac{3}{2} \frac{M^2}{r^2}  + \mathcal{O}((M/r)^3) \right).
\end{equation}
In all the calculations we will only need terms of $\mathcal{O}(r_0^{-2})$. Thus we can use $r_B(\bar{t}) = r_0 + d$ and $r_A(\bar{t})  = r_0$. 

\begin{equation}
	 \bar{l}_{AB} =  \int^{r_B(\bar{t})}_{r_A(\bar{t})}  dr \ \bar{g}_{rr}^\frac{1}{2}  = d + M \ln \left( \frac{r_0 + d}{r_0} \right) - \frac{3}{2} M^2 \left( \frac{1}{r_0 + d} - \frac{1}{r_0} \right)
\end{equation}
This can be fully simplified to $\mathcal{O}(r_0^{-2})$,

\begin{equation}
	 \bar{l}_{AB} =  d \left( 1 + \frac{M}{r_0} - \frac{1}{2} \frac{Md}{r_0^2} + \frac{3}{2} \frac{M^2}{r_0^2} \right).
\end{equation}

This is the proper length between $A$ and $B$ before shell crossing. 
The calculation for $\bar{\tau}_{AB}$ is much longer and so we do that separately in appendix \ref{prop-time}.   We refer the reader to Eq (\ref{tauini}) which shows that $\bar{\tau}_{AB} = 2 \bar{l}_{AB}$, thus proving our claim.  

\subsubsection{Proper length after shell crossing}

In this section we calculate the proper length between $A$ and $B$ after shell crossing. Integrating Eq (\ref{proplen}) with the trajectories for $r_B(t)$ and $r_A(t)$ gives 

\begin{equation}
	l_{AB} = r_B(t) - r_A(t) + (M - \delta M) \underbrace{\ln \left( \frac{r_B(t)}{r_A(t)} \right)}_{T9} - \underbrace{\frac{3}{2} (M - \delta M)^2 \left( \frac{1}{r_B(t)} - \frac{1}{r_A(t)} \right)}_{T10}
\end{equation}
Thus we see the only new thing to compute is $r_B(t) - r_A(t)$. Lets expand these terms individually, starting with $T9$

\begin{eqnarray}
	\ln \left( \frac{r_B(t)}{r_A(t)} \right) & = & \ln \left( \frac{(r_B - r_0) + r_0}{(r_A - r_0) + r_0} \right)	\nonumber	\\
	& = & \ln \left( \left( 1 + \frac{r_B - r_0}{r_0} \right) \left( 1 + \frac{r_A - r_0}{r_0} \right)^{-1} \right)	\nonumber	\\
	& = & \ln \left( 1 + \frac{r_B - r_A}{r_0} - \frac{(r_B - r_0)(r_A - r_0)}{r_0^2} - \left( \frac{r_A - r_0}{r_0} \right)^2 \right).
\end{eqnarray}
Note that $r_A - r_0 = \mathcal{O}(r_0^{-1})$ and we are only interested in terms up to $\mathcal{O}(r_0^{-2})$ thus we simplify the above equation to 
\begin{equation}
	\ln \left( \frac{r_B(t)}{r_A(t)} \right) = \ln \left( 1 + \frac{r_B - r_A}{r_0} \right).
\end{equation}
Now we look ahead to $r_B(t) - r_A(t)$ in Eq (\ref{rchange}). We will need to keep terms up to $\mathcal{O}(r_0^{-1})$ in $r_B(t) - r_A(t)$. Thus we expand the log to get

\begin{equation}
	\ln \left( \frac{r_B(t)}{r_A(t)} \right) = \frac{r_B - r_A}{r_0} - \frac{1}{2} \left( \frac{r_B - r_A}{r_0} \right)^2 = \frac{d}{r_0} + \frac{\delta M d}{r_0^2} - \frac{1}{2} \frac{d^2}{r_0^2} 
\end{equation}
Evaluating $T10$ to $\mathcal{O}(r_0^{-2})$ is straightforward

\begin{equation}
	- \frac{3}{2} (M - \delta M)^2 \left( \frac{1}{r_B(t)} - \frac{1}{r_A(t)} \right) = \frac{3}{2} \frac{(M -\delta M)^2d}{r_0^2}	\label{rsecond}
\end{equation}
Putting in the expansion of $T9$ and $T10$ into $l_{AB}$

\begin{equation}
	l_{AB} = r_B(t) - r_A(t) + \frac{(M - \delta M)d}{r_0} + \frac{\delta M (M - \delta M)d}{r_0^2} - \frac{1}{2} \frac{d^2(M- \delta M)}{r_0^2} + \frac{3}{2} \frac{(M - \delta M)^2d }{r_0^2}	\label{L1}
\end{equation}	

Now we can compute $r_B(t) - r_A(t)$. 
\begin{eqnarray}
	r_B(t) - r_A(t) & = & d - \frac{1}{2} \frac{M}{r_0^2} \bar{t}_c^2 - \left( \frac{M}{r_0^2} \bar{t}_c + \frac{\delta M}{r_0} + \frac{\delta M^2}{r_0^2} - \frac{\delta M d}{r_0^2} \right) (t - t_c) - \frac{1}{2} \frac{(M - \delta M)}{r_0^2} (t - t_c)^2 \label{rchange}
\end{eqnarray}

This expression contains both $t_c$ and $\bar{t}_c$. $t_c$ can be calculated using Eq (\ref{time}).

\begin{equation}
	\int^{t_c}_0 \ dt = \int^{r_B(t_c)}_{r_0} \frac{dr}{1 - \frac{2(M - \delta M)}{r}} \Rightarrow t_c = d \left( 1 + \frac{2(M - \delta M)}{r_0} + \mathcal{O}(r_0^{-2}) \right)	\label{tc}
\end{equation}

$\bar{t}_c$ can be calculated by using the IJC. In particular by demanding that the two induced metrics $h_{ab}(\bar{h}_{ab})$ corresponding to the two Schwarzschild metrics $g_{ab}(\bar{g}_{ab})$ are the same at the position of the shell. In general the induced metric $h_{ab}$ is given by

\begin{equation}
	h_{ab} = g_{\alpha \beta} e^\alpha_a e^\beta_b, \hspace{5mm} e^\alpha_a \equiv \frac{dx^\alpha}{dy^a}
\end{equation}

where $x \in \{t,r,\theta,\phi\}$ and $y\in \{t,\theta, \phi\}$. Thus for the two Schwarschild metrics in Eqs (\ref{SCH}, \ref{SCHa}) the corresponding induced metrics are

\begin{eqnarray}
	\bar{h}_{ab} & = & -\left( 1 - \frac{2M}{r} \right) d\bar{t}^2 + r^2 d \Omega_2^2	\nonumber	\\
	h_{ab} & = & -\left( 1 - \frac{2(M - \delta M)}{r} \right) dt^2 + r^2 d \Omega_2^2
\end{eqnarray}

Demanding that $\bar{h}_{ab}$ is equal to $h_{ab}$ at the position of the shell $r_0$ gives

\begin{equation}
	d\bar{t} = \sqrt{\frac{1 - \frac{2(M - \delta M)}{r_0}}{ 1- \frac{2M}{r_0}}} dt \Rightarrow d\bar{t} = \left( 1 + \mathcal{O}(r_0^{-1}) \right) dt.
\end{equation}

Thus we see that the time coordinates in both metrics are the same to leading order. We only need the leading order result for $\bar{t}_c$ as it will always multiply an acceleration (as the only thing that happens in the initial metric is acceleration). The acceleration is a $\mathcal{O}(r_0^{-2})$ quantity and therefore we can ignore any corrections and just use $\bar{t}_c = d$. 
Finally substituting Eqs (\ref{tc}, \ref{rchange}) into Eq (\ref{L1}) gives

\begin{eqnarray}
	l_{AB} & = & d + \frac{\delta M d}{r_0} + \frac{2\delta M(M - \delta M)d}{r_0^2} + \frac{\delta M^2}{r_0^2} d - \frac{1}{2} \frac{\delta M d^2}{r_0^2} + \frac{(M - \delta M)d}{r_0} + \frac{\delta M(M - \delta M) d}{r_0^2} - \frac{1}{2} \frac{d^2(M - \delta M)}{r_0^2} 	\nonumber	\\
	& + & \frac{3}{2} \frac{M^2 d}{r_0^2} - \frac{3 M \delta M d}{r_0^2} + \frac{3}{2} \frac{\delta M^2 d}{r_0^2}	\nonumber	\\
	& = & d + \frac{M d}{r_0} - \frac{1}{2} \frac{\delta M^2 d}{r_0^2} + \frac{3}{2} \frac{M^2 d}{r_0^2} -\frac{1}{2} \frac{Md^2}{r_0^2}	\label{lenfin}
\end{eqnarray}

\subsubsection{Longitudinal memory effect}

The memory effect in the longitudinal direction, $\Delta l_{AB}$ is the difference between $l_{AB}$ and $\bar{l}_{AB}$.

\begin{equation}
	\Delta l_{AB} \equiv l_{AB} - \bar{l}_{AB} = - \frac{\delta M^2 d}{r_0^2}.
\end{equation}
We note that there is no time dependent term in this expression. This follows from the fact that the velocity of both points is the same up to $\mathcal{O}(r_0^{-2})$.

Next, we will calculate the distance between $A$ and $C$. We will show that there will be a relative velocity between them and therefore a time dependent term at $\mathcal{O}(r_0^{-2})$. 

\subsection{Proper length change between $A$ and $C$}
\label{sec-lAC}
In this subsection we calculate the proper distance change between $A$ and $C$. First lets write down the trajectory equation for point $C$

\begin{eqnarray}
	r_{C}(\bar{t})  & = & d - \frac{1}{2} \bar{a}_{(2)}^{(C)}  \bar{t}^2 ,  \hspace{5mm} (\bar{t} \leq 0 )		\nonumber	\\
	r_{C}(t) & = &  d - \frac{1}{2} a_{(2)}^{(C)} t^2 - \left(v^{(C)}_{(1)} + v^{(C)}_{(2)} \right) t,  \hspace{5mm} (t >0) . 
\end{eqnarray}

The acceleration and velocities are defined below

\begin{eqnarray}
	\bar{a}^{(A)}_{(2)} & = & \bar{a}^{(B)}_{(2)} = \bar{a}^{(C)}_{(2)} = \frac{M}{r_0^2}	\nonumber	\\
	a^{(A)}_{(2)} & = & a^{(B)}_{(2)} = a^{(C)}_{(2)} = \frac{M - \delta M}{r_0^2}	\nonumber	\\
	v^{(A)}_{(1)} & = & v^{(B)}_{(1)} = v^{(C)}_{(1)}  =  \frac{\delta M}{r_0} 	\nonumber	\\
	v^{(A)}_{(2)} & = & v^{(C)}_{(2)} =  \frac{\delta M^2}{r_0^2}		
\end{eqnarray}
From the geometry in figure \ref{fig:1} we see that only $\theta$ and $r$ will change in the motion of $A$ and $C$. Thus we can drop the $\phi$ dependence from the start and write the integral for proper length in Eq (\ref{proplen}) as
\begin{equation}
	 l_{AC} = \int^{\theta_f}_{\theta_i} d \theta \ g_{\theta \theta}^\frac{1}{2} \left(  1 + \underbrace{(g_{rr}/g_{\theta \theta}) (dr/d\theta)^2}_{T8}  \right)^\frac{1}{2}.
\end{equation}
where $\theta_f = d/r_0$ and $\theta_i = 0$. To evaluate $T8$ we would need the geodesic equation for $r$ as function of $\theta$. Instead of evaluating it explicitly we note that $\frac{dr}{d \theta}$ will be of order $d$. Since $g_{\theta \theta}^{-1} = \frac{1}{r_0^2} + \mathcal{O}(r_0^{-3})$, to leading order we can ignore the effect of $T8$. Integrating the first term gives

\begin{equation}
	l_{AC} = r_C(t) (\theta_f - \theta_i) = r_C(t) \frac{d}{r_0}.
\end{equation}
Before shell crossing the proper length $\bar{l}_{AC} = d + \mathcal{O}(r_0^{-3})$. Whereas after shell crossing, there will be an additional velocity term in the proper length, $l_{AC}$, which will give a time dependent term at $\mathcal{O}(r_0^{-2})$

\begin{equation}
	l_{AC} = d - \frac{\delta M d}{r_0^2} t.
\end{equation}

The difference in proper lengths will be 

\begin{equation}
	\Delta l_{AC} \equiv l_{AC} - \bar{l}_{AC} = - \frac{\delta M d}{r_0^2} t.
\end{equation}

\section{Proper time for photon to go between $A$ and $B$ before shell crossing}
\label{prop-time}

From the metric, we can set $ds = 0$ for the photon to get 

\begin{equation}
	\int^{t + t_1}_{t} dt = \int^{r_B(t + t_1)}_{r_A(t)} \frac{dr}{\left( 1 - \frac{2M}{r} \right)}	\label{time}
\end{equation}

Integrating this gives the trajectory equation

\begin{equation}
	t_1 = r_B(t + t_1) - r_A(t) + 2M \ln \left( \frac{r_B(t + t_1) -2M}{r_A(t) - 2M} \right)	\label{B22}
\end{equation}

Now we can expand the log explicitly to see where terms of different orders are. 

\begin{equation}
	2M \ln \left( \frac{r_B(t + t_1) - 2M}{r_A(t) - 2M} \right) =  2M \left( \frac{r_B(t +t_1) - r_A(t)}{r_A(t)} \left( 1 + \frac{2M}{r_A(t)} \right) - \frac{1}{2} \left( \frac{r_B(t + t_1) - r_A(t)}{r_A(t)} \right)^2 \left( 1 -\frac{2M}{r_A(t)} \right)^{-2} \right)
\end{equation}

Since we only want to keep terms that are $\mathcal{O}(r_0^{-2})$ we can replace $\frac{M}{r_A(t)}$ with $\frac{M}{r_0}$. Eq (\ref{B22}) can now be written as

\begin{equation}
	t_1 = (r_B(t + t_1) - r_A(t)) \left( 1 + \frac{2M}{r_0} + \frac{4M^2 - M (r_B(t + t_1) - r_A(t))}{r_0^2} \right).
\end{equation}

Computing $r_B(t + t_1) + r_A(t)$ to $\mathcal{O}(r_0^{-2})$ gives.

\begin{equation}
	r_B(t + t_1) - r_A(t) = d - \frac{1}{2} \frac{M}{r_0^2}(t + d)^2 + \frac{1}{2} \frac{Mt^2}{r_0^2} = d \left( 1 - \frac{1}{2} \frac{Md}{r_0^2} - \frac{Mt}{r_0^2} \right)
\end{equation}

Therefore $t_1$ becomes

\begin{equation}
	t_1 = d \left( 1 + \frac{2M}{r_0} + \frac{4M^2}{r_0^2} - \frac{3}{2} \frac{Md}{r_0^2} - \frac{Mt}{r_0^2} \right)
\end{equation}

Similarly we can compute $t_2$.

\begin{equation}
	t_2 = (r_B(t + t_1) - r_A(t+t_1 +t_2)) \left( 1 + \frac{2M}{r_0} + \frac{4M^2 - M (r_B(t + t_1) - r_A(t+t_1+t_2))}{r_0^2} \right)	\label{A8}
\end{equation}

\begin{eqnarray}
	r_B(t + t_1) - r_A(t + t_1 + t_2) & = & d - \frac{1}{2} \frac{M}{r_0^2} (t + d)^2 + \frac{1}{2} \frac{M}{r_0^2} (t + 2d)^2	\nonumber	\\
	& = & d \left( 1 + \frac{Mt}{r_0^2} + \frac{3}{2} \frac{Md}{r_0^2} \right) 	\label{A9}
\end{eqnarray}

Substituting Eq (\ref{A9}) into Eq (\ref{A8}) gives 

\begin{eqnarray}
	t_2 & = & d \left( 1 + \frac{Mt}{r_0^2} + \frac{3}{2} \frac{Md}{r_0^2} \right) \left( 1 + \frac{2M}{r_0} + \frac{4M^2 - Md}{r_0^2} \right)\nonumber	\\
	& = & d \left( 1 + \frac{2M}{r_0} + \frac{4M^2}{r_0^2} + \frac{1}{2} \frac{Md}{r_0^2} + \frac{Mt}{r_0^2} \right)
\end{eqnarray}

The total coordinate time is $t_1 + t_2$,

\begin{equation}
	t_1 +t_2 = 2d \left(1 + \frac{2M}{r_0} + \frac{4M^2}{r_0^2} - \frac{1}{2} \frac{Md}{r_0^2}\right).
\end{equation}

Using this we can calculate the proper time

\begin{eqnarray}
	\tau & = &  \left( 1 - \frac{2M}{r} \right)^\frac{1}{2} (t_1 + t_2) 	\nonumber	\\
	& = & 2d \left( 1 - \frac{M}{r} - \frac{1}{2} \frac{M^2}{r_0^2} \right) \left( 1 + \frac{2M}{r_0} + \frac{4M^2}{r_0^2} - \frac{1}{2} \frac{Md}{r_0^2} \right)		\nonumber	\\
	&  = & 2d \left( 1  + \frac{M}{r_0} + \frac{3}{2} \frac{M^2}{r_0^2} - \frac{1}{2} \frac{Md}{r_0^2} \right).	\label{tauini}
\end{eqnarray}

\bibliography{all_active}

\end{document}